\documentclass[imslayout, preprint]{imsart}

\usepackage{amsmath}
\usepackage{amssymb}
\usepackage{graphics}
\usepackage{graphicx}

\usepackage{tikz}
\usepackage{collcell}

\usepackage{lineno,hyperref}

\usepackage{color}
\usepackage[utf8]{inputenc}
\usepackage[T1]{fontenc}

\startlocaldefs

\usepackage{colortbl}
\usepackage{pgfplots}
\usepackage{pgfplotstable}
\usepackage{booktabs}

\pgfplotstableset{
	/color cells/min/.initial=0,
	/color cells/max/.initial=1000,
	/color cells/textcolor/.initial=,
	color cells/.code={%
		\pgfqkeys{/color cells}{#1}%
		\pgfkeysalso{%
			postproc cell content/.code={%
				\begingroup
				\pgfkeysgetvalue{/pgfplots/table/@preprocessed cell content}\value
				\ifx\value\empty
				\endgroup
				\else
				\pgfmathfloatparsenumber{\value}%
				\pgfmathfloattofixed{\pgfmathresult}%
				\let\value=\pgfmathresult
				\pgfplotscolormapaccess
				[\pgfkeysvalueof{/color cells/min}:\pgfkeysvalueof{/color cells/max}]
				{\value}
				{\pgfkeysvalueof{/pgfplots/colormap name}}%
				\pgfkeysgetvalue{/pgfplots/table/@cell content}\typesetvalue
				\pgfkeysgetvalue{/color cells/textcolor}\textcolorvalue
				\toks0=\expandafter{\typesetvalue}%
				\xdef\temp{%
					\noexpand\pgfkeysalso{%
						@cell content={%
							\noexpand\cellcolor[rgb]{\pgfmathresult}%
							\noexpand\definecolor{mapped color}{rgb}{\pgfmathresult}%
							\ifx\textcolorvalue\empty
							\else
							\noexpand\color{\textcolorvalue}%
							\fi
							\the\toks0 %
						}%
					}%
				}%
				\endgroup
				\temp
				\fi
			}%
		}%
	}
}

\definecolor{verde}{rgb}{255,0.0,0.0}

\usepackage{soul}
\usepackage{longtable}
\usepackage{tabu}

\usepackage{adjustbox}
\usepackage{array}
\usepackage{multirow}

\newcolumntype{R}[2]{%
	>{\adjustbox{angle=#1,lap=\width-(#2)}\bgroup}%
	l%
	<{\egroup}%
}
\newcommand*\rot{\multicolumn{1}{R{90}{1em}}}

\usepackage{rotating}
\usepackage[authoryear]{natbib}

\endlocaldefs

\begin{document}

\begin{frontmatter}
\title{Brazilian Network of PhDs Working with Probability and Statistics}

 \begin{aug}
\author{ \fnms{Luciano} \snm{Digiampietri}\thanksref{LD}\ead[label=e1]{luciano.digiampietri@gmail.com}}
\author{ \fnms{Leandro} \snm{Rêgo}\thanksref{LR}\ead[label=e2]{leandrochavesrego@gmail.com}}
\author{ \fnms{Filipe} \snm{Costa de Souza}\thanksref{RO}\ead[label=e3]{filipecostadesouza@hotmail.com}}
\author{ \fnms{Raydonal} \snm{Ospina}\thanksref{RO}\ead[label=e4]{raydonal@de.ufpe.br}}
\and
\author{ \fnms{Jesús} \snm{Mena-Chalco}\thanksref{JM}\ead[label=e5]{jmenac@gmail.com}}

\affiliation[LD]{Universidade de São Paulo}
\affiliation[LR]{Universidade Federal do Ceará}
\affiliation[RO]{Universidade Federal de Pernambuco}
\affiliation[JM]{Universidade Federal do ABC}
 \address{Universidade Federal de Pernambuco\\
 Departamento de Estatística, \\Cidade Universitária,\\
 50740-540, Recife, PE, Brasil\\
  \printead{e4}\\
  \printead{e3}
}
 \end{aug}

\begin{abstract}
Statistical and probabilistic reasoning enlightens our judgments about uncertainty and the chance or beliefs on the occurrence of random events in everyday life. Therefore, there are scientists working with Probability and Statistics in various fields of knowledge, what favors the formation of scientific network collaborations of researchers with different backgrounds. Here, we propose to describe the Brazilian PhDs who work with probability and statistics. In particular, we analyze national and states collaboration networks of such researchers by calculating different metrics. We show that there is a greater concentration of nodes in and around the cites which host Probability and Statistics graduate programs.  Moreover, the states that host
P\&S Doctoral programs are the most central. We also observe a disparity in the size of the states networks. The clustering coefficient of the national network suggests that this network and regional differences especially with respect to states from South-east and North is not cohesive and, probably, it is in a maturing stage.
\end{abstract}


\end{frontmatter}

\section{Introduction}
\label{introduction}

Traditionally, academic collaboration is represented via co-authorship network (\cite{Glanzel2004}, \cite{Yoshikaneetal2004}, \cite{newman2001colaborationPart1}, \cite{Newman04Networks}, \cite{Neal201484}, and \cite{DeStefano2013370}). The use of co-authorship is especially useful for being a well-defined relationship,
because it is easy to obtain data, and it is possible to replicate or update such studies. However, \cite{kats1997what} argue that academic collaboration may be something much broader than co-authorship of scientific papers, including advisor-advisee relationships, and partnership in projects, classes, etc.
Moreover, \cite{melinPersson1996} warn that academic collaboration can also produce other pro\-ducts such as patents, or generate (in less than 5\% of the cases, as estimated by the authors) no tangible product at all.

Moreover, many performance studies used co-authorship metrics to explain academic performance using metrics such as number of articles written in English~\cite{Yousefi-Nooraie2008} g-index~\cite{Abbasi2011}, h-index~\cite{Cimenler2014} and research funds~\cite{Bellotti2012}. These works traditionally show that nodes position and/or types of relationship play an important role in academic productivity.

In this paper, we analyzed the collaboration network among PhDs working with Probability and Statistics (P\&S) in Brazil.  Some reasons motivate us to perform this work, for example, the majority of the studies about the relation between network and performance metrics are based on co-authorship. Here, we also analyze the academic social network of Brazilian states, where the ties are not limited to co-authorship, including participation in projects and advisor-advisee relationship. On the other hand, to our knowledge, there is no social network study devoted to analyze Brazilian researchers working in the P\&S field;

For the network design, the information contained in the Lattes platform (\url{ http://lattes.cnpq.br }) was considered, and the relationships analyzed include: coauthorship, participation in a research project and the advisor-advisee relationship. These relationships were examined along the last 35 years (from 1980 to 2014).  Different metrics were calculated for national and states collaboration networks of such researchers. We show that cities hosting graduate programs in P\&S aggregate the majority of PhDs in these fields and states networks are heterogeneous in their sizes.  Finally, the analysis of clustering coefficient of the national network indicates an immature stage of this network.

The remaining of this paper is organized as follows: Section ~\ref{relatedwork} summarizes related works regarding to: the P\&S area in Brazil, social network background and academic networks. Section~\ref{section:3} contains a brief description of Lattes platform from where the information of collaboration among Brazilian PhDs working with P\&S was collected. Section~\ref{methodology} describes the methodology used for the development of this work. In Section~\ref{results} the results are presented and discussed. Section~\ref{conclusions} contains the conclusions and directions for future work.

\section{Related Work}
\label{relatedwork}

There are different types of works related to this one, for example: about the domain, the theoretical background and related to the methodology. The first, discussed in Section \ref{sec2.1}, is composed of works which analyze the P\&S area in Brazil. The second, Section \ref{background} present the main concepts of social networks analysis; and the third contains works that use academic curricula in order to assess academic social networks.

\subsection{Probability and Statistics in Brazil}
\label{sec2.1}

Over the centuries, the probabilistic reasoning, and statistical and experimental methods are walking hand-by-hand with other scientific fields. It is almost impossible to imagine how the society could have evolved without that knowledge. From the medical industry to telecommunications. We are all surrounded by P\&S applied knowledge. Although many theoretical studies are dealing to improve P\&S methods, there are a much larger body of works applying it, and that is the main characteristic which makes P\&S area so especial, i.e., the massive interaction with other fields. 

Therefore, it is plausible to imagine that there is a rich collaboration environment among those working with P\&S. However, we still have little bibliometric information about this community, especially in Brazil. In this paper, we explore the scholarly networks of PhDs working with P\&S in Brazil, considering the academic relationships from 1980 to 2014. To better understand the history of probability and statistics in Brazil we recommend the following readings: \cite{SENRA2008,SENRA2009,ara2012}.

According to the Brazilian Ministry of Education website
(e-MEC\footnote{\url{http://emec.mec.gov.br/} (accessed on 11/11/2015).})  there are 81 undergraduate courses in statistics (at  University of São Paulo, there is also a BA in Applied and Computational Mathematics with an emphasis on economic statistics) in Brazil. In this context, it is noteworthy that, as stated by \cite{ara2012}, knowledge about statistics permeate virtually all undergraduate courses in Brazil, and for being an evidence-based science, assists in the scientific development of different areas.

The P\&S Graduate Courses are evaluated by the Mathematics,
Probability and Statistics Committee from CAPES\footnote{Brazilian Coordination for the Improvement of Higher Education.}. The result of the last assessment showed  that there are nine Statistics Graduate Courses (one of them is a Mathematics and Statistics Graduate Course, and other is an Applied Mathematics and Statistics Graduate Course). From these nine courses, six have a PhD Program\footnote{\url{http://www.avaliacaotrienal2013.capes.gov.br} (accessed on 11/11/2015).}. Moreover, from 2010 to 2012 these programs graduate 79 PhDs, with the largest contribution being made by USP (University of São Paulo) with 36 defenses. Regarding the CNPq
research productivity fellowship (a fellowship targeted at researchers who stand out among their peers, enhancing their scientific production according to normative criteria established by CNPq), there are currently 70 fellows in payroll, 38 level 2, 8 level 1D, 4 level 1C, 15 level 1B and 5 level 1A\footnote
{\url{http://plsql1.cnpq.br/divulg/} (accessed on 01/14/2015).}, being 1A the highest level and the 2 the lowest one.

\cite{ara2012}, through a sample survey, describe the profile of professors of undergraduate Statistics courses at Brazilian Public Universities, especially regarding academic education. The authors found that most professors (63\%) are not statisticians. Besides, among those with master's degree, only 31\% has a master degree in statistics, and among the PhDs, only 20\% have the PhD degree in statistics. The North and Northeast regions have the
highest percentage of professors with an undergraduate degree in statistics (60\%), on the other hand, in the South, only 12\% have an undergraduate degree in statistics. The Southern region has the highest percentage of professors with Doctoral degrees (83\%) and the North has the lowest percentage (38\%). But among doctors, the scenario is reversed: the South has the lowest percentage of professors with a PhD degree in statistics (15\%), while the northern region has
the highest percentage (33\%).

\subsection{Social Network Background}
\label{background}

According to \cite{Easley:2010}, a network or a graph is way to represent relationships among a collection of elements, named nodes or vertices. Formally, a network is a pair $(N, M)$, where $N = \{1, 2,\ldots, n\}$ is a collection of elements or simply the finite set of nodes, and $M$ is a $n\times n$ matrix, where $m_{ij}$ represents the relationship between node $i$ and node $j$. If nodes $i$ and $j$ in $N$ are related, then we say that there is a link (or an arrow, or an edge, or a tie) between them.  Depending on some characteristics of $M$, the network could be classified in different manners. When, for all $i$ and $j$ in $N$, $m_{ij} = m_{ji}$ the graph is said to be direct, otherwise, it is called undirected. Undirected graph happens when the relationship is not reciprocal, e.g, an advisor-advised relationship. Moreover, when all values in $M$ are taken from $\{0, 1\}$ the graph is said to be unweighted, where $m_{ij} = 1$ express that nodes $i$ and $j$ are related and $m_{ij} = 0$ indicates the absence of relationship. On the other hand, if the values of $M$ could assume more than two values (expressing the intensity of the relationship) then the network is said to be weighted \cite{Jackson:2008}. 

As stated by \cite{Digiampietri2011_ijac} the characterization of a network to be direct (or not) or to be unweighted (or not) depends on the type of relationships analyzed. In academic networks, the relationships among nodes traditionally represent co-authorship or others tips of academic interaction such as advisor-advised relationship, partnership in projects etc., so the values in $M$ are always non-negative integers. 

\cite{DeStefano2011} report that some academic connections are clearly direct or weighted, most of academic networks are treated as undirected and unweighted, especially when they also involve co-authorship or multiple relationships. This is justified because for many researches, the main goal is to identify (and understand) the relationship between academics, institutions and countries. 
In this context, to transform a weighted graph to its unweighted version, one shall simply to set all values in $M$ that are greater than zero to one; and to transform a direct graph to its undirected form, one shall set $m_{ij}=m_{ji}=1$ every time that one pair of nodes $i$ and $j$ have $m_{ij}\neq m_{ji}.$   

So, in what follows, we will discuss some concepts and metrics related to unweighted and undirect networks as could also be seen in \cite{Jackson:2008} and \cite{Mena2012}:

\begin{itemize}
	\item Total number of links: a link between two nodes indicates that they maintained a relationship in the network. Therefore, the total number of links indicates the total number of connections in the network during the analyzed period.
	
	\item Connected network: a network is connected if all its nodes can reach one another by a sequence of ties.
	
	\item Component of a network: a component of a network $(N, M)$ is a connected subnetwork $(N', M')$ where $N'$ is a nonempty subset of $N$ and $M'$ is a submatrix of $M$ such that if $i \in N'$ and  
	$m_{ij}=1$ in $M$, than $j \in N'$ and $m_{ij}=1$ is preserved in $M'$.

	\item Size of component: is the total number of nodes in a given component. The Biggest component in the network is called the giant component 
	
	\item Maximum clique size: correspond to the maximum subset of vertices in which everybody is related with each other.
	
	\item Degree of a node (or Degree Centrality): is the number of ties involving a given node. Nodes with degree equal to zero are called isolated.
	
	\item Average degree: is the sum of the degree of each node in the network divided by the total number of nodes. 
	\item E-I index: is a segregation metric \cite{Bojanowski-Corten2014}, proposed by \cite{Krackhardt-Stern1988}, to evaluate the relationship between external and internal links in a network. By simplicity, suppose $N$ was partitioned into two non-empty disjoint groups (one called internal group ($IG$) and the other called the external group ($EG$)). Let $EL$ be the total number of links between nodes from $IG$ and the nodes from $EG$ (i.e., we only count a tie if one node is from $IG$ and the other from $EG$); let $IL$ be the total number of links only between nodes from $IG$ and, finally, let $T = EL + IL$. Then the E-I index ($EI$) is formulated as $EI = (EL-IL)/T$. This index ranges from $-1$ (expressing that all links are internals) to $+1$ (expressing that all links are externals). 
	
	\item Density: indicates the ratio between the number of edges in the network and the maximum number of possible edges, i.e., it indicates how close the network is to be complete. Density equals to zero means that there are no edges in the graph, on the other hand, density equal to one means that all nodes are connected to each other.
	
	\item Diameter: is the maximum distance (or path) between any two nodes in the network. However, if the network is disconnected, then, the diameter of the network will be the biggest one among the diameters of each network component. So, the diameter could vary from one (in the best case scenario) to $\#N-1$ (in the worst case scenario).    
	
	\item Closeness Centrality: is the inverse of the average distance between a given node and all others nodes in its component. 
	
	\item Betweenness centrality: is the average proportion of short paths that a given node lied on . Therefore, this metric indicates how important a node is to link other vertices.
	
	\item Eigenvector centrality: express the importance of a node in the network based on the importance of its neighbors; i.e., as stated by \cite{bonacich2001}, this metric is relevant when nodes' status is determined by their neighbors.
	
	\item   Centralization: for each centrality metric (e.g. degree, betweenness, closeness and eigenvector), it indicates (based in a specific centrality measure) how central is the most central node in the network. These metrics are based on the sum of the differences between the most central vertex and all other vertices, divided by the theoretical maximum sum of differences. For more details see \cite{FREEMAN1978215}.
	
	\item Cluster coefficient: express the proportion of the vertices of a given node who also have a link between them \cite{latapy2008basic}. The average cluster coefficient of a network is the mean value of the cluster coefficient of its nodes. Therefore, the cluster coefficient measures the transitivity of the relationships in the graph. The value 1 (one) means that the relationships are all transitive, while the value 0 means that the relationships are all intransitive. 
\end{itemize}

\subsection{Academic Social Networks}

By using social network analysis, researchers may understand and evaluate academic interactions in many ways. According to \cite{melinPersson1996}, using co-authorship we are able to study collaboration among researchers (the traditional co-authorship network), or to study institutional collaboration (when we analyze co-authorship among different institutions based on the author's professional address), or even international collaboration (co-author partnership among countries). Nevertheless, the authors also recognized that academic collaboration is something larger than co-authorship, once it can lead to other types of products and knowledge.

\cite{mahlctpersson2000} studied co-authorship and citation networks from two departments at different Swedish universities between 1986 and 1996. To analyze the co-authorship network the authors used the concept of socio-bibliometric maps, where nodes (authors) were labeled according to some status such as gender, academic degree, etc. and links (relationship) besides indicating co-authorship could also highlight if the persons had an advisor-student relationship. Among the results, the authors found that the most productive authors were PhDs, and they were surrounded by less productive ones, who were, mostly, their students.

\cite{Yousefi-Nooraie2008} analyzed the co-authorship networks of three Iranian Medical academic research centers to study its scientific productivity (articles written in English). As a result, authors found that centers with denser and more decentralized networks, and that are also more open to outside connections had better scientific outcomes.

\cite{Bellotti2012} studied how variations in network measures (in micro, i.e., collaboration between scientists; in macro level, i.e., between institutions; and in meso level, i.e., micro and macro metrics combined) could explain variations in the total money that an Italian Physicist receives to fund his/her research. As a result, the author inferred that researchers that collaborate with many different Physicists (i.e. that change partners over the years) tend to get more money. This characteristic was even more important than working in a big university or having many connections in the network.

Concerning to social network studies dealing (in some manner) with the P\&S community, we shall highlight the work from \cite{Baccini2009}. The authors studied editorial politics of Statistics \& Probability journals creating a network where two journals are linked if they share a same editor in their boards. Moreover, the editorial proximity of two journals could be valued by the strength of the tie. The resulting network was very compact, which could be seen, according to the authors, as evidence of a common perspective about appropriate investigation methods and theoretical development in the domain of Probability and Statistics.

\cite{DeStefano2011} critically discussed some issues in the analysis of co-authorship networks such as: data collection, network boundary setting, relational data matrix definition, data analysis and interpretation of results. Furthermore, authors illustrated their argumentation using real data based on researchers involved in four disciplines (Physics, Engineering, Arts \& Humanities and Economics \& Statistics) at the Italian university of Salerno.

\cite{DeStefano2013} aimed to compare co-authorship network results of Italian academic statisticians using three data sources (Web of Science, Current Index to Statistics and nationally funded research projects). As a result, authors observed the small-world structure of the networks and for some statistic subfields they also found evidences that the authors seem to behave as if they are guided by a scale-free distribution. Furthermore, the general idea of positive association between statisticians’ performance (h-index) and their central positions in the network was confirmed. However, some results may depend on the Bibliographic archives source.

As done by \cite{Abbasi2011} and \cite{Cimenler2014}, \cite{Bordons2015} run a Poisson regression model to studied the relationship between the research performance (g-index) of scien\-tists and his/her posi\-tion in co-author\-ship network. Moreover, the authors analyzed three co-author ship networks (Nano\-science, Pharmacology and Statistics) in Spain during 2006 to 2008, to understand trends in each one of the fields. As a result, they found that Statistic Network was less dense, less connected and more fragmented than the others. The degree centrality and the strength of links were positive related with the g-index in all three fields; however, the benefits (in terms of g-index) from the author position in the network were smaller in the Statistics field. 

\cite{Said2010} proposed a model of preferential attachment in co-author\-ship networks and used it to predict emerging scientific subfields over time. They argued that the process of one actor attaching to another and strengthening the tie over time is a stochastic random process based on the distributions of tie-strength and clique size among authors. Thus, they used empirical data of statisticians working in prominent American Universities, focusing on the biopharmaceutical subfield, to estimate these distributions.

In Brazil, there are studies about co-authorship in several areas of knowledge. \cite{mena-chalco2009} developed a software named scriptLattes that extracts and analyzes data from the Curriculum Lattes, and it became an important and useful tool for those interested in academic network and bibliometric analysis. Using the scriptLattes, \cite{MenaChalco2014} were able to evaluate over one million curriculums of Brazilian researches. \cite{andretta2012} studied the scientific production of graduate programs in Information Science in Brazil, analyzing issues such as the profile of the production, productivity, and scientific collaboration, highlighting the characteristics of each Brazilian region. \cite{Andrettaetal2012} repeated the same study focusing on the State of São Paulo. \cite{ALVES2014} evaluated the profile of CNPq’s research productivity fellows in Chemistry in Brazil. \cite{costa2013} investigated the scientific collaboration among the Brazilian northeast researchers working in biotechnology. These authors also identified which are the main universities in the region connected with foreign centers. \cite{NASCIMENTO2011} studied the scientific production networks among graduate programs in Accounting in Brazil.

\section{Brazilian Probability and Statistics Dataset}
\label{section:3}

Each research can register in his/her Lattes curriculum from zero to six expertise areas. For this, the areas in the Lattes Platform are represented by four levels of hierarchy, namely: major knowledge area; area; subarea; and specialty.

There are nine major knowledge areas that can be chosen by the researcher, as follows: Exact and Earth Sciences; Biological Sciences; Engineering; Health Sciences; Agricultural Sciences; Applied Social Sciences; Human Sciences; Linguistics, Letters and Arts; and Other. The Exact and Earth Sciences major area is divided into eight areas, namely: Mathematics; Probability and Statistics; Computer Science; Astronomy; Physics; Chemistry; Geosciences; and Oceanography.

The P\&S area, in turn, is divided into three subareas:
Statistics; Probability; and Applied Statistics and Probability. The Statistics subarea is divided into eight specialties: Data Analysis; Multivariate Analysis; Fundamentals of Statistics; Inference in Stochastic Processes; Non-Parametric Inference; Parametric Inference; Design of Experiments; Regression and Correlation. The Probability subarea is divided into six specialties: Stochastic Analysis; Special Stochastic Processes; Markov Processes; Limit Theorems; General Theory and Probability Foundations; General Theory and Stochastic
Processes. The Applied Probability and Statistics subarea is a specialty in itself.

Some authors, such as \cite{arruda2009}, argue that this division imposed by Lattes curricula structure is not clear for certain research purposes. In addition, certain knowledge areas (and their subsequent levels in the hierarchy) include issues related to P\&S, for example, there is a specialty named Methods and Mathematical Models, Econometric and Statistics, in the Quantitative Methods in Economics subarea, in the Economics area that belongs to the major knowledge area of Applied Social Sciences. This sequence can easily be confused with the Probability and Applied Statistics subarea/specialty. Besides, authors can also include other subareas in the Lattes curricula if they do not find an adequate one to describe their research.

\section{Method}
\label{methodology}

The methodology used in this paper was organized in four activities: data gathering; sample selection; relevant information extraction; calculation of metrics. 

In the \emph{data gathering} activity, it was used the XML raw file from 3.2 millions of curricula from Lattes Platform\footnote{\url{http://lattes.cnpq.br/}}.
These curricula were downloaded by the researchers from the Social Network Analysis and Scientometrics
Group\footnote{\url{http://dgp.cnpq.br/dgp/espelhogrupo/9125239221851493}} in July 2013. This group aims to study the characteristics of the entire Brazilian scientific production and developed a methodology to obtain and organize the curricula files from Lattes Platform. More details see \cite{Digiampietri2014a} and \cite{MenaChalco2014}.

In the \emph{sample selection} activity, from the 3.2 million curricula, were select for this study the ones that satisfy three criteria: curricula from PhDs (first criterion) that contain the ``Probability and Statistics'' value in the field ``Areas of Expertise'' (second criterion) and which professional address is in Brazil (third criterion). This activity identified 2,373 curricula. 

In the \emph{relevant information extraction} step, the following information from the Lattes curricula were extracted and organized:
professional address (for geolocation of the curricula); expertise areas; relationships among curricula (the Lattes Curricula have explicit relationships information of coauthors, advisors, advisees, members of a scientific project, etc.).

In this data set of study, we observe that 2,147 (91.61\%) researchers were born in Brazil and the others are from 41 different nations; especially from Peru (48), Argentina (23), France (11) and Cuba, Colombia and the United States (10 PhDs each). Moreover, 2,226 PhDs (93.81\%) have Brazilian nationality. Table \ref{tbl:PhdRegion}  shows the distribution of PhDs with Brazilian nationality by Region. One can clearly see that the foreigners PhDs traditionally work in the Southeast region (69.39\%).
\begin{table}[hbt]
	\centering
\caption{Distribution of PhDs working with P\&S  by nationality and Brazilian region.}
\renewcommand{\tabcolsep}{0.9pc} 
\renewcommand{\arraystretch}{1.2} 
\begin{tabular}{cccc} \hline
	Region    & Brazilian & Foreign & Total \\ \hline
	North     & 68        & 4             & 72    \\
	Northeast & 338       & 12            & 350   \\
	Central-West   & 179       & 13            & 192   \\
	Southeast & 1276      & 102           & 1378  \\
	South     & 365       & 16            & 381   \\ \hline
	{\bf	Brazil }   &  2226      & 147           & 2373 \\ \hline 
\end{tabular}

\label{tbl:PhdRegion}
\end{table}

Table \ref{tbl:Phdchar} summarizes some characteristics of PhDs working with P\&S by Region. There we can see that the Southeast is the region that concentrates the majority (over 58\%) of PhDs working in P\&S in Brazil; on the other hand, the North region concentrates only 3\% of them. Regarding to advisorship activities, the PhDs form the Midwest had advised on average 6.64 graduation students; PhDs form Northeast had advised on average 4.21 Scientific initiation students; PhDs from South had advised on average 3.24 specialization students and 4.99 master dissertations; while Southeast PhDs had advised on average 4.62 doctoral theses. 

\begin{table}[hbt]
	\centering
{\scriptsize	
	\caption{Some informations of PhDs researchers working with P\&S by Brazilian region}
	\label{tbl:Phdchar}
	\renewcommand{\tabcolsep}{0.4pc} 
	\renewcommand{\arraystretch}{1.2} 
	
	\begin{tabular}{llm{1.4cm}m{1.2cm}m{1.3cm}m{1.1cm}m{1.3cm}m{1.1cm}} \hline
	Region    & PhDs \% & Graduation students advised & Scientific initiation students advised & Speciali\-za\-tion students advised & Master students advised & PhD \ \ \ \ stu\-dents ad\-vised & Others tips of advisorship \\ 
	\hline 
	North     & 3.03    & 4.21                       & 3.51                                   & 1.67                           & 2.82                   & 0.22                & 0.54                      \\
	Northeast & 14.75   & 3.60                        & 4.21                                   & 1.58                           & 3.84                   & 0.60                 & 1.79                      \\
	Central-West   & 8.09    & 6.64                       & 3.73                                   & 1.66                           & 3.04                   & 0.38                & 1.79                      \\
	Southeast & 58.07   & 3.89                       & 2.84                                   & 1.08                           & 4.62                   & 4.62                & 2.18                      \\
	South     & 16.06   & 5.07                       & 3.59                                   & 3.24                           & 4.99                   & 1.09                & 2.46                      \\
	Brazil    & 100     & 4.27                       & 3.26                                   & 1.56                           & 4.38                   & 1.13                & 2.08                      \\ \hline 
\end{tabular}
} 

\end{table}

 In addition to these explicit relationships, we also used the algorithm presented in~\cite{100371} for the identification of coauthorship relationships that were not explicitly present in the curricula (the absence of this information on the curricula occurs, typically, due to the lack of standardization in the filling of the name of the authors in the publications' registers). All these relationships (explicit or not) were used in the production of the academic social networks. After that, 29 social networks (graphs) were produced: one composed of the 2,373 researchers from the sample, and 27 additional networks composed only of researchers from each one of the Brazilian States and the Federal District and one network considering each state as a node.

In the \emph{calculation of metrics} activity, we measured the social networks structural metrics (See ~\cite{Wasserman1994}). These metrics aim to explain some characteristics from the networks to allow their understanding and comparison.

\section{Results }
\label{results}

Based on the relationships from the 2,373 curricula, 29 academic social networks
were produced. These networks will be presented and analyzed in this section.

\subsection{Data Description}

We first describe characteristics of the PhDs in our sample according to their location (Table~\ref{tbl:general}) and expertise (Table~\ref{tbl:expertise}).

The Brazilian government estimates its population at 202 million people (The Brazilian Institute of Geography and Statistics - IBGE\footnote{\url{http://www.ibge.gov.br/home/estatistica/populacao/estimativa2014/}}). More than half (55\%) of the Brazilian population is located in the five most populous states (respectively, SP, MG, RJ, BA and RS). These five states contains 66\% of the PhDs working with P\&S. The variables \emph{Percentage of PhDs Working with P\&S} and \emph{Average time since PhD graduation} are highly correlated (the value of the Pearson correlation is 0.544 with p-value $<$ 0.05). The most populous state (SP) has the highest \emph{Average time since PhD graduation} (13.9 years) and the three less populous have the lowest values for these variable (AP, AC and RR, with 4, 6.3 and 6.5 years, respectively). The number of \emph{PhD working with P\&S in Brazil per million people} ranges from 1 (in AP) to 31 (in DF). On average, there are 12 PhDs working with P\&S in Brazil per million people. The highest concentration of PhDs per million people occurs in the Federal District (DF) which contains universities and federal agencies that employ many of these PhDs.

\begin{table}[!hbt]
	\caption{\label{tbl:general} Some informations of the PhDs working with P\&S by Brazilian states.}
\centering
\footnotesize
\renewcommand{\tabcolsep}{0.45pc} 
\renewcommand{\arraystretch}{1.2} 
   \begin{tabular}{lrccccc}
     \hline
		      &            &               & PhDs    & PhDs       & Average    						& PhDs   \\
		State & Population & Pop. \%       & working & working    & time since           	& working with  \\
		      &            &               & with    & with    & PhD & P\&S per   \\
					& 					 &               & P\&S    & P\&S \% & graduation     & million people \\
		\hline
AC & 790,101 & 0.39\% & 4 & 0.17\% & 6.3 & 5 \\
AL & 3,321,730 & 1.64\% & 17 & 0.72\% & 10.8 & 5 \\
AM & 3,873,743 & 1.91\% & 20 & 0.84\% & 11.4 & 5 \\
AP & 750,912 & 0.37\% & 1 & 0.04\% & 4.0 & 1 \\
BA & 15,126,371 & 7.46\% & 75 & 3.16\% & 9.3 & 5 \\
CE & 8,842,791 & 4.36\% & 52 & 2.19\% & 9.8 & 6 \\
DF & 2,852,372 & 1.41\% & 87 & 3.67\% & 12.3 & 31 \\
ES & 3,885,049 & 1.92\% & 37 & 1.56\% & 8.1 & 10 \\
GO & 6,523,222 & 3.22\% & 44 & 1.85\% & 8.3 & 7 \\
MA & 6,850,884 & 3.38\% & 12 & 0.51\% & 8.6 & 2 \\
MG & 20,734,097 & 10.22\% & 281 & 11.84\% & 10.1 & 14 \\
MS & 2,619,657 & 1.29\% & 30 & 1.26\% & 10.2 & 11 \\
MT & 3,224,357 & 1.59\% & 31 & 1.31\% & 8.1 & 10 \\
PA & 8,073,924 & 3.98\% & 32 & 1.35\% & 8.8 & 4 \\
PB & 3,943,885 & 1.94\% & 50 & 2.11\% & 8.5 & 13 \\
PE & 9,319,347 & 4.60\% & 76 & 3.20\% & 11.8 & 8 \\
PI & 3,194,718 & 1.58\% & 5 & 0.21\% & 12.2 & 2 \\
PR & 11,081,692 & 5.46\% & 145 & 6.11\% & 9.9 & 13 \\
RJ & 16,461,173 & 8.12\% & 378 & 15.93\% & 11.8 & 23 \\
RN & 3,408,510 & 1.68\% & 50 & 2.11\% & 9.4 & 15 \\
RO & 1,748,531 & 0.86\% & 8 & 0.34\% & 7.8 & 5 \\
RR & 496,936 & 0.25\% & 2 & 0.08\% & 6.5 & 4 \\
RS & 11,207,274 & 5.53\% & 160 & 6.74\% & 11.0 & 14 \\
SC & 6,727,148 & 3.32\% & 76 & 3.20\% & 11.9 & 11 \\
SE & 2,219,574 & 1.09\% & 13 & 0.55\% & 10.3 & 6 \\
SP & 44,035,304 & 21.71\% & 682 & 28.74\% & 13.9 & 15 \\
TO & 1,496,880 & 0.74\% & 5 & 0.21\% & 8.2 & 3 \\
\hline
\textbf{Brazil} & 202,810,182 & 100.00\% & 2,373 & 100.00\% & 11.5 & 12 \\
\hline
   \end{tabular}

\end{table}

Regarding the expertise of the PhDs working with P\&S, in Table~\ref{tbl:expertise}, we can see that two subareas concentrate the great majority of such PhDs: Applied Probability and Statistics (41.84\%) and Statistics (38.31\%). Therefore, approximately 8 out of 10 PhDs working with P\&S areas, work in at least one of these subareas. In third place, there are 7.82\% of such PhDs working in the Probability subarea. This suggests that the great majority of the PhDs working with P\&S develop more applied than theoretical researches.

\begin{table}[!hbt]
\centering
\caption{Distribution of PhDs according with their expertise.}

\renewcommand{\tabcolsep}{0.5pc} 
\renewcommand{\arraystretch}{1.2} 
   \begin{tabular}{lc}
     \hline
		{Subarea} & {Percentage} \\
		     \hline
Applied Probability and Statistics & 41.84\% \\
Statistics & 38.31\% \\
Probability & 7.82\% \\
Biostatistics & 0.95\% \\
Time Series & 0.44\% \\
Applied Statistics & 0.36\% \\
Spatial Statistics & 0.36\% \\
Multivariate Statistics & 0.28\% \\
Bayesian Inference & 0.28\% \\
Econometrics & 0.24\% \\
Experimental Statistics & 0.24\% \\
Stochastic Processes & 0.24\% \\
Other & 8.65\% \\
\hline
\end{tabular}
\label{tbl:expertise}
\end{table}

\subsection{The Academic Social Network of Researchers in Probability and Statistics}

\begin{figure}[!hbt]
	\centering
	\includegraphics[width=.95\textwidth]{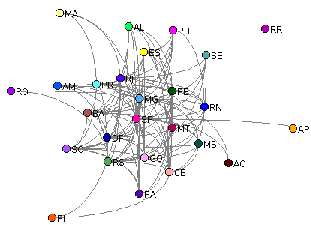}
	\caption{P\&S collaboration network --  Brazilian states network}
	\label{fig:estaares}
\end{figure}

We will start our analysis focusing in the Brazilian state network, in which each state (including the Federal District-DF) is treated as a node, as shown in Figure \ref{fig:estaares}. The nodes were disposed in the graph in order to arrange those with higher centrality measures in the center and the ones with lower measures in the periphery. In this context, four central states should be highlighted: SP, MG, RJ and PE. According to the CAPES's triennial\footnote{\url{http://www.avaliacaotrienal2013.capes.gov.br/relatorios-de-avaliacao}} assessment report 2013, those are the unique Brazilian states with Statistic doctoral programs (SP--USP, UNICAMP and UFSCAR; MG--UFMG; RJ--UFRJ; PE--UFPE).

We can see in Figure~\ref{fig:estaares} that the network is disconnected because of RR. Furthermore, its diameter is equal to 3 and its average path length is 1.63. These two metrics indicate the traditional small word idea ( see \cite{Travers1969} and \cite{milgram1967}).

The average degree is 11.48 and the average cluster coefficient 0.59. \cite{Jackson:2008}  pointed that a feature of social networks is that they tend to have high cluster coefficient when compared to random networks. For example, a random network with 27 nodes and an average degree of 11.48 has a cluster coefficient of about 0.43 (11.48/27), which corroborates this trend. Other metrics are summarized in Table \ref{tbl:globalmetric1}.  

Table~\ref{tbl:globalmetric2} exposes four centrality metrics (degree, betweenness, closeness and eigenvector) of each state. Clearly, the most central states are from the Southeast region, specially, SP, MG and RJ. Moreover, when one observes the centralization metrics in Table~\ref{tbl:globalmetric1}, particularly the centralization degree and centralization eigenvector, it is possible to understand the importance of SP as the central and most important and connected vertex of the network. On the other hand, the states from the north region were the less central.

\begin{table}[!hbt]
	\centering
	\caption{Global metric of P\&S collaboration network --  Brazilian states network}
	\label{tbl:globalmetric1}
	\renewcommand{\tabcolsep}{0.5pc} 
	\renewcommand{\arraystretch}{1.1} 
	\begin{tabular}{ll} \hline
		Metric & Value \\ \hline
		Clique Number            & 8.00  \\
		Average Path Length     & 1.63 \\
		Clustering Coefficient   & 0.59 \\
		Centralization degree    & 0.52 \\
		Centralization closeness & 0.23 \\
		Centrelization evcent    & 0.52 \\
		Diameter                  & 3.00 \\
		Graph density            & 0.41 \\ \hline
	\end{tabular}
\end{table}

\begin{table}[!hbt]
	\centering
	\caption{Centrality metrics of P\&S collaboration network by Brazilian states}
	\label{tbl:globalmetric2}
		\renewcommand{\tabcolsep}{0.5pc} 
		\renewcommand{\arraystretch}{1.1} 
	\begin{tabular}{ccccc} \hline
		State & Betweenness & Closeness & Degree   & Eigenvalue \\ \hline
		AC     & 0.000       & 0.013     &  4  & 0.155      \\
		AL     & 0.002       & 0.014     & 10 & 0.499      \\
		AM     & 0.000       & 0.014     &  7  & 0.325      \\
		AP     & 0.000       & 0.013     &  3  & 0.079      \\
		BA     & 0.013       & 0.016     & 16 & 0.781      \\
		CE     & 0.042       & 0.016     & 16 & 0.728      \\
		DF     & 0.006       & 0.015     & 13 & 0.644      \\
		ES     & 0.007       & 0.015     & 13 & 0.648      \\
		GO     & 0.004       & 0.014     & 10 & 0.461      \\
		MA     & 0.000       & 0.013     &  5  & 0.207      \\
		MG     & 0.121       & 0.018     & 23 & 0.966      \\
		MS     & 0.003       & 0.015     & 11 & 0.544      \\
		MT     & 0.003       & 0.014     & 11 & 0.538      \\
		PA     & 0.003       & 0.014     & 10 & 0.469      \\
		PB     & 0.001       & 0.014     &  9  & 0.448      \\
		PE     & 0.027       & 0.016     & 18 & 0.835      \\
		PI     & 0.000       & 0.013     &  4  & 0.123      \\
		PR     & 0.032       & 0.016     & 17 & 0.785      \\
		RJ     & 0.092       & 0.017     & 20 & 0.840      \\
		RN     & 0.009       & 0.015     & 13 & 0.621      \\
		RO     & 0.000       & 0.013     &  4  & 0.143      \\
		RR     & 0.000       & 0.001     &  2  & 0.000      \\
		RS     & 0.058       & 0.017     & 19 & 0.850      \\
		SC     & 0.004       & 0.015     & 12 & 0.599      \\
		SE     & 0.000       & 0.014     &  7  & 0.326      \\
		SP     & 0.199       & 0.019     & 25 & 1.000      \\
		TO     & 0.001       & 0.014     &  8  & 0.353    \\ \hline 
	\end{tabular}
\end{table}

\begin{figure}[!hbt]
\centering
        \includegraphics[width=1.0\textwidth]{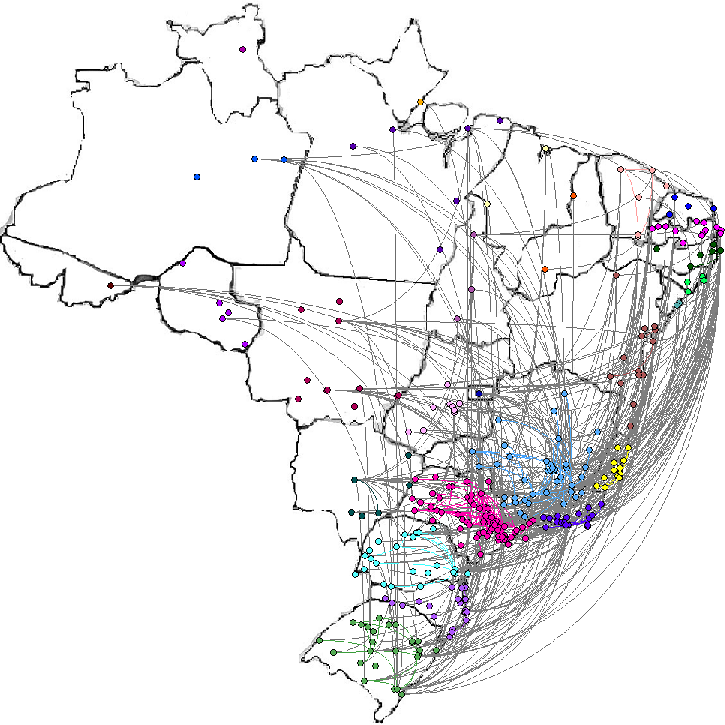}\\
    \includegraphics[width=0.55\textwidth]{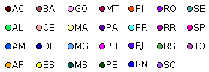}
    \caption{Probability and Statistics Collaboration Network - Brazilian cities.}
    \label{fig:cities}
\end{figure}

In the network from Figure~\ref{fig:cities}, each node represents a Brazilian city (in which there is at least one PhD working with P\&S). The edges between cities indicate there is at least one  academic collaboration between the PhDs from these cities. Edges between cities from the same state are colored, and the ones between cities from different states are gray. It is possible to observe a concentration of nodes (cities) in the states from the South and Southeast regions. On the other hand, in the North and Central-West regions there are few cities with PhDs working with P\&S.

\begin{figure}[!htb]
	\centering
	\includegraphics[width=1.0\textwidth]{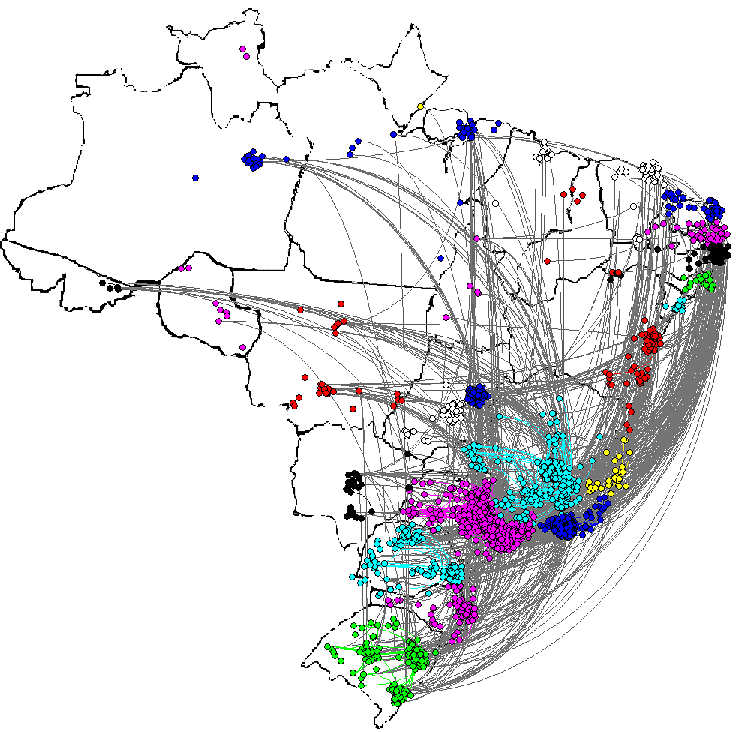}
	\\
    \includegraphics[width=0.55\textwidth]{figure01.png}
	\caption{Probability and Statistics Researchers Collaboration  Network distributed according to their professional address.}
	\label{fig:network}
\end{figure}

In the network from Figure~\ref{fig:network}, each node represents one PhD working with P\&S. Each node was positioned in the Brazilian Map close to its professional address (in order to minimize overlays, the nodes were not positioned exactly over their professional address). The network's edges correspond to the relationship between two PhDs (nodes). In this figure, it is possible to observe the concentration of nodes in the states' capitals and, as seen in Figure~\ref{fig:cities}, a concentration of nodes in the states from the South and Southeast regions, followed by the states in the Northeast region. It is worth to note that there is a greater concentration of nodes in and around the cities which host P\&S graduate programs, namely: Brasília, Belo Horizonte, Belém, Recife, Rio de Janeiro, Natal, São Paulo, Campinas and São Carlos.

\begin{figure}[!hbt]
\centering
   \includegraphics[width=1.0\textwidth]{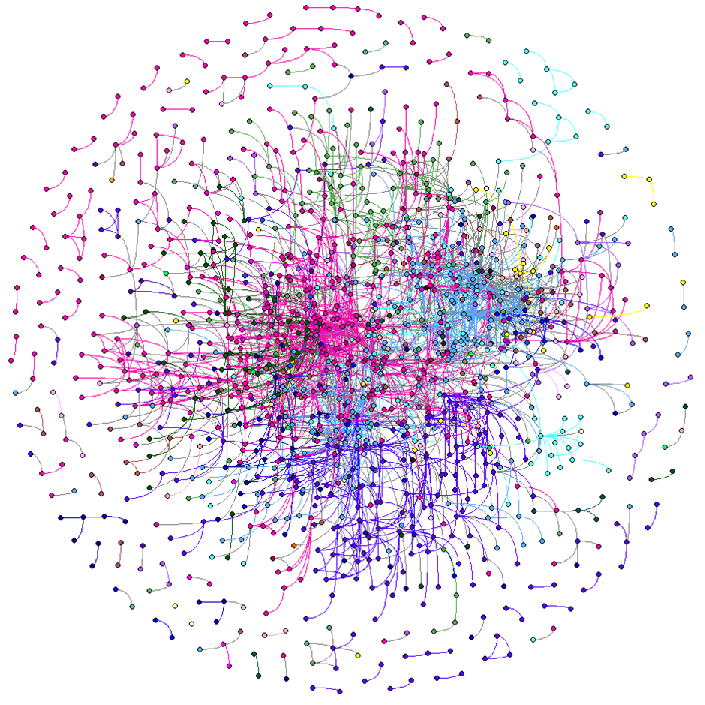}\\
    \includegraphics[width=0.55\textwidth]{figure01.png}
   \caption{Probability and Statistics Researchers Collaboration Network - reorganized.}
   \label{fig:reorganized}
\end{figure}

Figure~\ref{fig:reorganized} aims to clarify the relationships among the PhDs working with P\&S. The network presented in this figure corresponds to two modifications applied in the network from Figure~\ref{fig:network}: nodes with degree zero were excluded; and the nodes were reorganized to approximate the nodes that are related and to move away the nodes that are not related. It is possible to observe in the center of Figure~\ref{fig:reorganized} the giant component. It is composed by the majority of the nodes of this network. Surrounding this component, there are dozens of smaller components, most composed only by two or three nodes.

Table~\ref{tbl:components} presents the distribution of the number of connected components from Figure~\ref{fig:reorganized} according to their size. As observed, the network contains a giant component with 1,197 nodes and several small components (composed by two to seven nodes). It is important to observe that almost more than one thousand PhDs (about 44\% of PhDs studied in this paper) do not have any relationship with other PhD from the sample. These PhDs were not plotted in the network presented in Figure~\ref{fig:reorganized}. 

\begin{table}[hbt]
	\caption{Number and size of the connected components.}
	\label{tbl:components}
		\renewcommand{\tabcolsep}{0.5pc} 
		\renewcommand{\arraystretch}{1.1} 
   \begin{tabular}{cc}
     \hline
		{Component Size} & {Number of Components} \\
		     \hline
1 & 1036 \\
2 & 54 \\
3 & 10 \\
4 & 7 \\
5 & 3 \\
6 & 1 \\
7 & 1 \\
1197 & 1 \\
\hline
\end{tabular}
\end{table}

\begin{figure}[hbt]
\centering

    \includegraphics[width=0.98\textwidth]{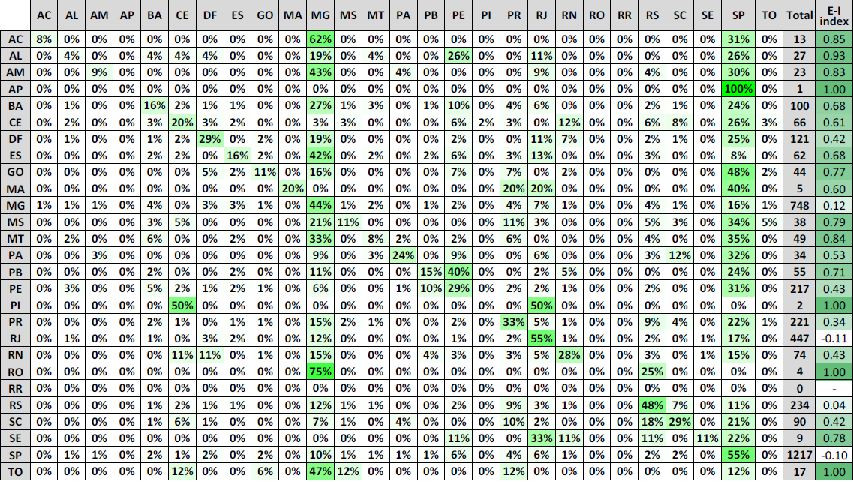}
        \caption{Percentage of relationships among states.}
    \label{tbl:edges}
\end{figure}

Figure~\ref{tbl:edges} presents the percentage of edges between the PhDs from each state. Each line in the table sums 100\%, corresponding to  all the relationships from the respective state. The penultimate column presents the total number of links from the respective state. Furthermore, the last column presents the E-I index. 

\begin{figure}[hbt]
\centering
	\includegraphics[width=.85\textwidth]{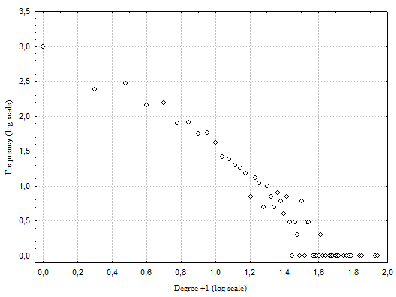}
	\caption{Log-log degree distribution -- Brazilian Network} 
			\label{fig:degree1}
\end{figure}

\begin{figure}[hbt]
\centering
	\includegraphics[angle=90, width=.85\textwidth]{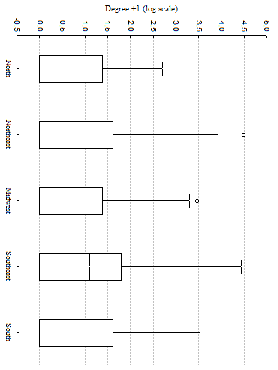}
		\caption{Distribution of ${\rm log}({\rm degree}+1)$ by Brazilian region}
		\label{fig:degree2}
\end{figure}

Figure~\ref{fig:degree1} presents the degree distribution of the Brazilian network. Over 40\% of the nodes have degree zero, and less than 10\% of the researchers have degree equal to ten or higher, with only one PhD from PE with degree equal to 87, indicating the rich-get-richer ideia (\cite{Easley:2010}). Additionally, still based on the degree distribution, we use the Kruskal-Wallis test to investigate regional degree differences. 
Figure~\ref{fig:degree2} shows the distribution of the ln(degree +1) for each Brazilian region. With a $p$-value of 0.0001 (KW-H(4, 2,373)=22.6793) there are statistical evidences against the null hypothesis, i.e., we can detect differences in degree distribution by region, specially between North and Southeast regions.

Assuming each Brazilian state as a group\footnote{We may think in a particular state as the $IG$, and the others as the $EG$, for example.} in the network, we are able to calculate the E-I index for each one of them\footnote{It is worth noting that, as the researchers from Roraima (RR) does not have internal or external links in the network, it was not possible to calculate the E-I index for RR.}. Most of the states had positive E-I index, with the exception of Rio de Janeiro (RJ) and São Paulo (SP). This result seems natural since these states are the most developed ones in Brazil, and researchers in these locations can find partners more easily in their own state. On the other hand, states from the north and northeast seem to have a greater degree of external dependence because the E-I index in these states are close to 1, or even equal to 1. 

Still according to Figure~\ref{tbl:edges}, the cell's background is colored according to its value \textcolor{blue}{(higher values implies in a higher green  tonality)}. It is worth to highlight three different information from this table. The first indicates the importance of some states concerning to relationships (the states whose columns have more green cells), especially São Paulo (SP) and Minas Gerais (MG), followed by Pernambuco (PE) and Rio de Janeiro (RJ). The second is the percentage of self-relationships (relationships between nodes from the same state), in this criteria, the most important states are: São Paulo (SP), Rio de Janeiro(RJ), Rio Grande do Sul (RS) and Minas Gerais (MG). At last, it is possible to observe that most of the remaining green cells corresponds to relationships between states geographically close, for example, 50\% of the relationships involving PhDs (nodes) from Piauí (PI) occur with PhDs from Ceará (CE); 20\% of the relationships involving nodes from Maranhão (MA) occur with nodes from Pará (PA); and 26\% from the relationships involving nodes from Alagoas (AL) and Paraíba (PB) occurs with PhDs from Pernambuco(PE) (neighboring states).

Table~\ref{tbl:metrics} contains the network metrics that were measured in both the national and the states level. The networks are sorted according to the number of nodes, and the ten biggest state networks are highlighted. Following, some characteristics of these networks will be discussed, focusing on the biggest networks. It is possible to observe a disparity in the size of the networks: the five smallest networks contains, together, less than 0.75\% of the total number of nodes. On the other hand, the São Paulo network contains more than 28\% of the nodes, and, adding the nodes from Rio de Janeiro and Minas Gerais they represent more than half of the total number of PhDs working in Brazil with P\&S. Together, the ten biggest state networks contain 85\% of the nodes.

\begin{sidewaystable}[htb]
	\caption{Networks' metrics.}
	\label{tbl:metrics}

		\renewcommand{\tabcolsep}{0.6pc} 
			\renewcommand{\arraystretch}{1.3} 
{\tiny
\begin{tabular}{@{}lllllllllllllllll@{}} \hline 
\rot{\shortstack[l]{State}}  
&\rot{\shortstack[l]{Nodes}}  
& \rot{\shortstack[l]{Nodes with degree \\ greater than zero}} 
& \rot{\shortstack[l]{Edges}}    
& \rot{\shortstack[l]{Nodes in the \\ giant component}} 
& \rot{\shortstack[l]{ \% of nodes in the \\ giant component}}  
& \rot{\shortstack[l]{Density}} 
& \rot{\shortstack[l]{Average \\Degree}} 
& \rot{\shortstack[l]{Clustering \\Coefficient}} 
& \rot{\shortstack[l]{Degree \\Centralization}} 
& \rot{\shortstack[l]{Closeness \\Centralization}} 
& \rot{\shortstack[l]{Betweeness \\Centralization}} 
& \rot{\shortstack[l]{Diameter}} 
& \rot{\shortstack[l]{Maximum \\Clique Size} }
& \rot{\shortstack[l]{Average Path\\ Length}} 
& \rot{\shortstack[l]{Eigenvector \\Centralization}}        \\ \hline
AP     & 1                                   & 0     & 0                            & 1                                  & -       & -              & -                      & -                     & -                        & -                         & -        & 0                   & 1                   & -                         & -     \\
RR     & 2                                   & 0     & 0                            & 1                                  & -       & -              & -                      & -                     & -                        & -                         & -        & 0                   & 1                   & -                         & -     \\
AC     & 4                                   & 2     & 1                            & 2                                  & 100.00   & 1.000          & 1.000                  & -                     & 0.167                    & 0.139                     & 0.000    & 1                   & 2                   & 1.000                     & 1.000 \\
PI     & 5                                   & 0     & 0                            & 1                                  & -       & -              & -                      & -                     & -                        & -                         & 0.000    & 0                   & 1                   & -                         & -     \\
TO     & 5                                   & 0     & 0                            & 1                                  & -       & -              & -                      & -                     & -                        & -                         & 0.000    & 0                   & 1                   & -                         & -     \\
RO     & 8                                   & 0     & 0                            & 1                                  & -       & -              & -                      & -                     & -                        & -                         & 0.000    & 0                   & 1                   & -                         & -     \\
MA     & 12                                  & 2     & 1                            & 2                                  & 100.00   & 1.000          & 1.000                  & -                     & 0.076                    & 0.014                     & 0.000    & 1                   & 2                   & 1.000                     & 1.000 \\
SE     & 13                                  & 2     & 1                            & 2                                  & 100.00   & 1.000          & 1.000                  & -                     & 0.071                    & 0.012                     & 0.000    & 1                   & 2                   & 1.000                     & 1.000 \\
AL     & 17                                  & 2     & 1                            & 2                                  & 100.00   & 1.000          & 1.000                  & -                     & 0.055                    & 0.007                     & 0.000    & 1                   & 2                   & 1.000                     & 1.000 \\
AM     & 20                                  & 3     & 2                            & 3                                  & 100.00   & 0.667          & 1.333                  & 0.000                 & 0.095                    & 0.010                     & 0.006    & 2                   & 2                   & 1.333                     & 0.977 \\
MS     & 30                                  & 6     & 4                            & 4                                  & 66.70   & 0.267          & 1.333                  & 0.000                 & 0.094                    & 0.007                     & 0.007    & 2                   & 2                   & 1.429                     & 0.974 \\
MT     & 31                                  & 6     & 4                            & 4                                  & 66.70   & 0.267          & 1.333                  & 0.000                 & 0.058                    & 0.006                     & 0.004    & 3                   & 2                   & 1.571                     & 0.957 \\
PA     & 32                                  & 10    & 8                            & 4                                  & 40.00   & 0.178          & 1.600                  & 0.500                 & 0.048                    & 0.005                     & 0.004    & 3                   & 3                   & 1.417                     & 0.967 \\
ES     & 37                                  & 16    & 10                           & 5                                  & 31.30   & 0.083          & 1.250                  & 0.000                 & 0.096                    & 0.005                     & 0.009    & 2                   & 2                   & 1.410                     & 0.971 \\
GO     & 44                                  & 9     & 5                            & 3                                  & 33.30   & 0.139          & 1.111                  & 0.000                 & 0.041                    & 0.002                     & 0.001    & 2                   & 2                   & 1.167                     & 0.990 \\
PB     & 50                                  & 9     & 8                            & 7                                  & 77.80   & 0.222          & 1.778                  & 0.231                 & 0.096                    & 0.005                     & 0.011    & 3                   & 3                   & 1.818                     & 0.964 \\
RN     & 50                                  & 19    & 21                           & 8                                  & 42.10   & 0.123          & 2.211                  & 0.500                 & 0.064                    & 0.005                     & 0.005    & 5                   & 3                   & 1.915                     & 0.924 \\
\rowcolor{black!5} \cellcolor{white}
CE     & 52                                  & 13    & 13                           & 5                                  & 38.50   & {\bf 0.167}          & 2.000                  & {\bf 0.833}                 & 0.049                    & 0.003                     & 0.003    & 3                   & 4                   & {\bf 1.390}                     & 0.960 \\
\rowcolor{black!5} \cellcolor{white}
BA     & 75                                  & 23    & 16                           & 6                                  & 26.10   & 0.063          & 1.391                  & 0.300                 & {\bf 0.035}                    & 0.002                     & {\bf 0.002}    & 4                   & 3                   & 1.679                     & 0.976 \\
\rowcolor{black!5} \cellcolor{white}
SC     & 76                                  & 29    & 26                           & 9                                  & 31.00   & 0.064          & 1.793                  & 0.255                 & 0.084                    & 0.002                     & 0.008    & 3                   & 3                   & 1.783                     & 0.967 \\
\rowcolor{black!5} \cellcolor{white}
PE     & 76                                  & 42    & 62                           & 33                                 & 78.60   & 0.072          & 2.952                  & 0.295                 & 0.165                    & 0.011                     & 0.094    & 6                   & 4                   & 2.809                     & 0.910 \\
\rowcolor{black!5} \cellcolor{white}
DF     & 87                                  & 31    & 35                           & 19                                 & 61.30   & 0.075          & 2.258                  & 0.270                 & 0.107                    & 0.005                     & 0.026    & 5                   & 4                   & 2.370                     & 0.939 \\
\rowcolor{black!5} \cellcolor{white}
PR     & 145                                 & 60    & 73                           & 41                                 & 68.30  & 0.041          & 2.433                  & 0.309                 & 0.083                    & 0.004                     & 0.041    & 8                   & 5                   & 3.818                     & 0.960 \\
\rowcolor{black!5} \cellcolor{white}
RS     & 160                                 & 76    & 112                          & 63                                 & 82.90   & 0.039          & 2.947                  & 0.307                 & 0.054                    & 0.005                     & 0.043    & 11                  & 4                   & 4.398                     & 0.893 \\
\rowcolor{black!5} \cellcolor{white}
MG     & 281                                 & 172   & 330                          & 151                                & 87.80   & 0.022          & 3.837                  & 0.261                 & 0.084                    & 0.004                     & 0.077    & 11                  & 6                   & 4.399                     & 0.952 \\
\rowcolor{black!5} \cellcolor{white}
RJ     & 378                                 & 183   & 248                          & 135                                & 73.80   & 0.015          & 2.710                  & 0.218                 & 0.039                    & 0.002                     & 0.046    & 12                  & 6                   & 4.893                     & 0.977 \\
\rowcolor{black!5} \cellcolor{white}
SP     & 682                                 & 384   & 671                          & 291                                & 75.80   & 0.009          & 3.495                  & 0.239                 & 0.037                    & 0.001                     & 0.042    & 18                  & 5                   & 5.714                     & 0.972 \\ \hline
\rowcolor{black!10} \cellcolor{white}
Brazil & 2373                                & 1391  & 2791                         & 1197                               & 86.10   & 0.003          & 4.013                  & 0.155                 & 0.036                    & 0.001                     & 0.032    & 17                  & 6                   & 5.547                     & 0.963\\ \hline
\end{tabular}
}
\end{sidewaystable}

The third column from Table~\ref{tbl:metrics} contains the number of nodes with degree greater than zero (i.e., the nodes that have at least one relationship) in each network. In the national network, only 1,391 from the 2,373 PhDs have at least one relationship. The calculation of the remainder metrics presented in this table used only the nodes with degree greater than zero. The fourth column presents the number of edges in each network. The national network contains 2,791 edges. It is worth to observe that, despite being the third greatest network, the Minas Gerais network is the second one in the number of edges. This characteristic influences several metrics as will be presented as follows.

As explained in Section \ref{background}  the size of the giant component corresponds to the number of nodes in the biggest connected component in each network.  In this work the percentage of nodes in the giant component was calculated considering only the nodes with degree greater than zero. Giant components with a high percentage of the network's nodes are considered a positive aspect in a social network. This indicates that a significant number of individuals belongs to the main flow of information/knowledge in the network. In the national network, 86.1\% of the nodes with degree greater than zero are in the giant component. Among the biggest state networks, stands out the Minas Gerais network with 87.8\% of its nodes in its giant component.  
Based on Density measure all networks assessed in this paper are far to be compĺete.The density of the national network is 0.003 (i.e., there are only 1/333 of the number of possible edges for this network). The density of Ceará network is 0.167 (one sixth of the maximum number of edges for this network).

In Table~\ref{tbl:metrics} for all states, the  average degree  is greater than or equal to one.  In the national network, the average degree is 4.01. In the biggest state networks, this value ranges from 1.39 (Bahia network) to 3.84 (Minas Gerais network). Still according to Table \ref{tbl:metrics} the Ceará network (CE) stands out for having a high clustering coefficient (0.833). This value indicates that the Ceará network is cohesive. The clustering coefficient of the national network is 0.155 which suggests that this network is not cohesive and, probably, it is in a maturing stage.

The centrality metrics indicate how influential are the nodes in a network.
High centrality values used to be viewed with caution in social networks, because they indicate that one individual is very important for the network. Thus, its eventual absence could imply in a great prejudice for the network.  In the national network, the degree and closeness centralities have low value (0.036 and 0.001, respectively). In the state networks, the highest centrality value occurs in the Pernambuco network, and the lowest in the Bahia network (degree centrality) and São Paulo (closeness centrality).

The diameter from Ceará and Santa Catarina networks is only three. The São Paulo network's diameter is 18. The diameter from the national network is 17, this value lower than the one from São Paulo network means in the national network there are shorter paths linking two persons from the same state (e.g., there are two PhDs in São Paulo which are not linked and do not have any common neighbor in the São Paulo network but they have a common neighbor in Mato Grosso do Sul network). The diameter in a social network used to be associated with the maximum time required to an information to be propagated to all the individuals from the network. Thus, lower values from this metric allows a faster information (or knowledge) propagation.

 A clique with a great amount of people typically represents a cohesive group of PhDs (for example, a research group) that works in some particular area/subarea of expertise. The size of the maximum clique in the national, Rio de Janeiro, and Minas Gerais networks is six. On the other hand, in Ceará and Santa Catarina networks this value is only three.

The average path length corresponds to the average distance of the shortest path between all pairs of individuals in a connected component. This metric is also related to information (or knowledge) propagation speed in the network. In the national network, in average, the path between two PhDs is composed of 5.55 persons. In the Ceará network, this value is only 1.39. In the São Paulo network, the average path length is 5.71.

\section{Final Remarks}
\label{conclusions}

In this paper, we performed a social network analysis of PhDs working in the field of Probability and Statistics in Brazil, where ties represent either co-authorship, participation in joint project or the advisor-advisee relationship. Particularly, 29 networks were analyzed one for each state, one considering each state as a node and one for the whole country.  As a result, the first network had small world characteristic, and the most central nodes were the states that host P\&S doctoral programs. Regional differences were also detected. The biggest networks were form southeast and the smaller were from the north region. The same characteristic was observed with respect to the degree distribution. The national network shows that there is a greater concentration of nodes in and around cities having graduate programs in Probability and Statistics, which is also reflected in the size of the state networks. The clustering coefficient of the national network suggests that this community is not cohesive and, probably, it is in a maturing stage. Moreover, the E-I index indicated that states from the north and northeast have a greater dependence on collaboration with researchers from other states.    

For further studies, we intend to investigate how network metrics can impact in productivity measures, which will allow us to use more sophisticated statistical methods in the (social) network analysis, as done by \cite{Abbasi2011}, \cite{Cimenler2014}, \cite{Bellotti2012}, \cite{deArruda:2013} and \cite{peron2012}.

\bibliographystyle{chicago}


\end{document}